\documentstyle[aps,prl]{revtex}
\begin{document}
\draft
\twocolumn[\hsize\textwidth\columnwidth\hsize\csname @twocolumnfalse\endcsname
\title{Roughening Transition in a One-Dimensional Growth Process}
\author{U. Alon$^{1}$, M.R. Evans$^{2}$, H. Hinrichsen$^{1}$,
	and D. Mukamel$^{1}$\\[-3mm]$ $}
\address{$^{1}$
	Department of Physics of Complex Systems,
        Weizmann Institute, Rehovot 76100, Israel}
\address{$^{2}$
	Department of Physics and Astronomy, University of Edinburgh,
	Mayfield Road, Edinburgh EH9 35Z, U.K.}
\date{Submitted to Physical Review Letters, 06 December 1995}
\maketitle
\begin{abstract}
A class of nonequilibrium models with short-range interactions
and sequential updates is presented. The models
describe one dimensional
growth processes which display a roughening transition
between a smooth and a rough phase.
This transition is accompanied by spontaneous symmetry breaking, which
is described by an order parameter whose dynamics is non-conserving.
Some aspects of models in this class are related to directed
percolation in 1+1 dimensions, although unlike directed percolation
 the models have
no absorbing states. Scaling relations are derived and
compared with Monte Carlo simulations.
\end{abstract}
\pacs{PACS numbers: 05.70Fh,05.70Jk,05.70Ln}]
\renewcommand{\theparagraph}{\Alph{paragraph}}
\input epsf

\def\FigOne{
\begin{figure}[h]
        \epsfxsize=90mm
        \centerline{ \epsffile{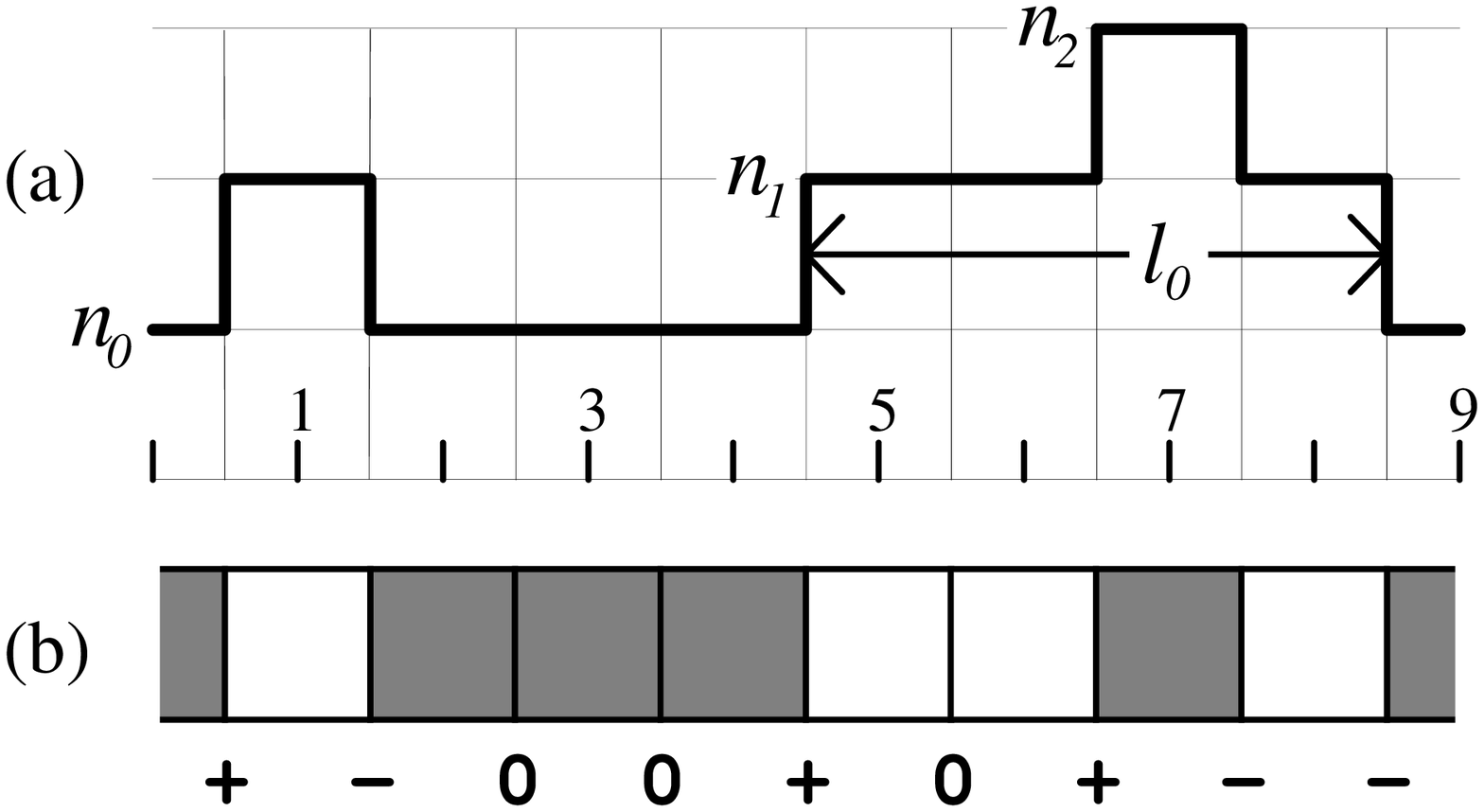}  }
\vspace{-3mm}
   {\small  FIG. 1. \
(a) Typical configuration of the interface. $n_k$ is the fraction of
   sites at height $k$ above the minimal height in the configuration
   (here $n_0=n_1=\frac49$, $n_2=\frac19$). The average island size grown
   on top of level $k$ is $l_k$.
(b) Mapping of the configuration of Fig.1a to the charged-particle
   representation, along with a site coloring, as described
   in Sec.~\ref{sec:ssb}.}
\end{figure}
\noindent
}
\def\FigTwo{
\begin{figure}[h]
        \epsfxsize=90mm
        \centerline{ \epsffile{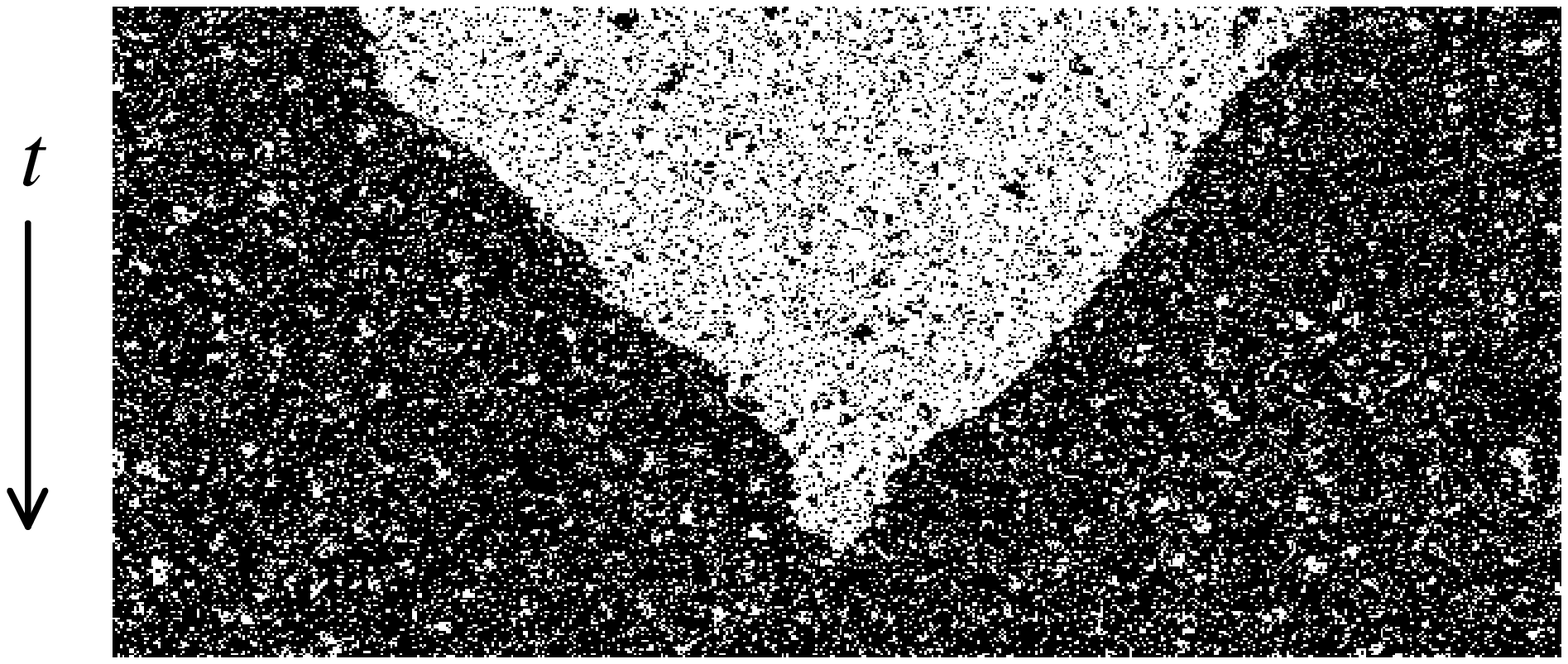}  }
    \vskip -20mm
{\small FIG. 2. \ Monte-Carlo simulation of the RSOS model for a
system of size 600 at $q=0.130$ where $q_c=0.189 \pm 0.002$. Each configuration
is a row of pixels, with sites at even and odd heights represented by
black and white pixels, respectively. Configurations at intervals of
7 moves per site (sweeps) are shown (time advances downwards) up to
2100 sweeps. At an early time, a large island of size 400 is introduced.
The island shrinks and disappears, illustrating the mechanism that
insures long-ranged order at $q<q_c$. Note that small islands and
islands within islands are continually generated by fluctuations,
and are washed away.}
\end{figure}
\noindent
}
\def\FigThree{
\begin{figure}[h]
        \epsfxsize=90mm
        \centerline{ \epsffile{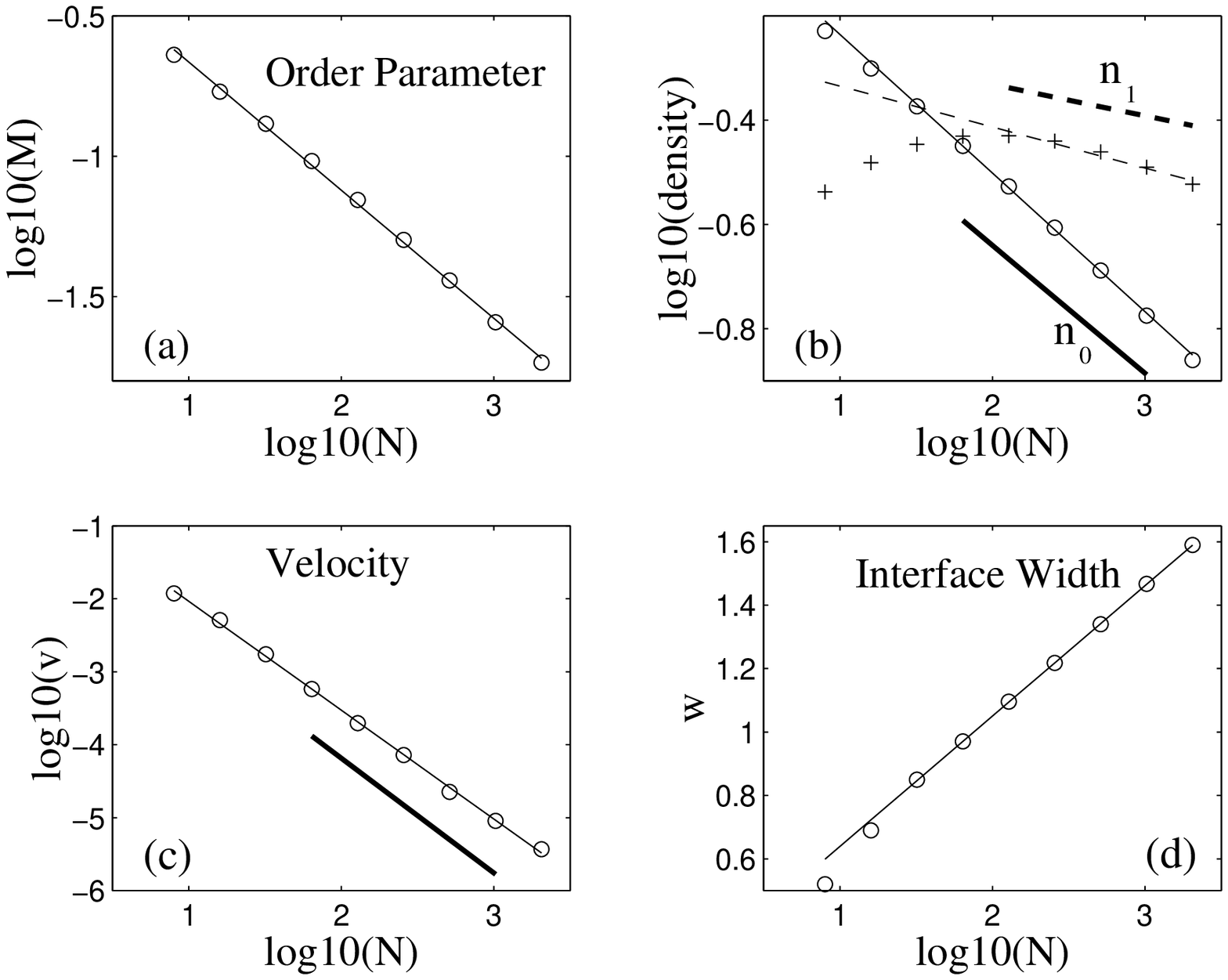}  }
    \vspace{5mm}
{\small FIG. 3. \ Monte-Carlo simulation results for the unrestricted
growth model. The critical behavior of a quantity $F \sim (q-q_c)^{\alpha}$
is estimated by finite size scaling, measuring $F$ in systems of size $N$
at the critical point $q=q_c$, and using $F(N) \sim N^{\alpha/\nu_{\perp}}$,
where $\nu_{\perp}\approx 1.10$ is the critical exponent associated with the
divergence of the correlation length (see Sec. \ref{sec:scaling}).
Systems of sizes $N=2^m, m=3\ldots10$ were studied at $q=q_c=0.233$
with $2^{23} \approx 8 \cdot 10^6$ moves per site.
(a) Order parameter $M$.
(b) Density of exposed sites on the lowest
   exposed level, $n_0$, and the next level $n_1$. The bold lines have
   the slopes expected from the scaling arguments
   (Eqs. \ref {eq n0}, \ref {eq n1}).
(c) Interface velocity $v$. The bold line has the slope expected from
   the scaling arguments (Eq. \ref {eq v}).
(d) Interface width, $w$,
   suggesting  $w \sim \log(N)$ (note that this graph is
   semi-logarithmic). }
\end{figure}
\noindent
}

The morphology of growing interfaces has attracted much interest
in recent years \cite{growth review}. Many growth processes of
two-dimensional surfaces exhibit a roughening transition, from
a smooth phase with finite width to a rough one with diverging width.
A question of interest is whether a one dimensional
(1d) growing interface  with short-range interactions and
unbounded noise, can exhibit a roughening transition \cite {libschaber}.
It is well known that in thermal equilibrium
no such phase transition can take place as 1d interfaces are always rough.
Growth processes far from equilibrium are, however, less restrictive
and the question of whether they are capable of exhibiting a roughening
transition in 1d is more subtle. Most growth
processes, such as those described by the KPZ equation \cite{KPZ},
are always rough in 1d. A class of 1d models, which have a maximal
velocity by which the uppermost point of the surface can propagate, has
been shown to display a roughening transition \cite {savit,kertesz}.
The existence of a maximal velocity in these models is due to the use
of parallel updates, and the smooth phase disappears if sequential
(continuous time) updates are used. Sequential updates are a more
adequate description of systems where different particles are not
exactly synchronized.
The question of whether a sequential update growth process is capable
of exhibiting a transition from a smooth to a rough phase is still open.

A related and more general question is that of spontaneous symmetry
breaking (SSB) and long range order in 1d systems \cite{gacs}.  Recently a
nonequilibrium 1d model with short-range interactions and unbounded
noise which exhibits SSB in the thermodynamic limit was presented
\cite{mukamel,mukamel2}. The model belongs to a class
of driven diffusive systems, in which charges
of two kinds are injected at both ends of a 1d lattice and are biased to
move in opposite directions. The microscopic rules are symmetric
under space and charge inversion. However, this symmetry is broken
in the steady state of the system where the
currents of the two charge species are different.
In this model SSB is a result of the {\it conserved} dynamics of the order
parameter in the bulk (charges are not created
or annihilated, except at the boundaries), and the existence of
{\it open} boundaries (two endpoints) at which the dynamics is different from
that of the rest of the system \cite{gel}. These two features create favorable
conditions for SSB. The conserved dynamics slows down the evolution
of the system; moreover flips between one broken symmetry phase
to another can originate only at the two boundary points, where the
order parameter is not conserved.  Simple
attempts to modify the model such that either one or both of these
features are eliminated results in symmetric steady states with no SSB.
It would therefore be of interest to examine the possibility of SSB
in 1d systems under more general conditions,
namely in homogeneous systems with
periodic boundary conditions and order parameters
with non conserving dynamics.

Finally, phase transitions in homogeneous
nonequilibrium 1d systems have usually been observed
in the past in systems which have absorbing states
(a set of states from which the system can not escape).
The canonical example is the `dry' state
below the percolation threshold in directed percolation models
\cite {directed percolation,domany-kinzel,kinzel}.
Thus it is of interest to find 1d
models with no absorbing states that display a phase transition.

In this Letter we present a class of nonequilibrium models with
short-range interactions and sequential updates,
which addresses the three questions posed above: The models describe
1d growth processes which display a roughening transition
between smooth and rough phases. This transition is accompanied by
SSB associated with a non-conserved order parameter in a homogeneous
system with periodic boundary conditions.
The models supply a robust local
mechanism for eliminating islands of minority phases generated by
fluctuations in the bulk of the majority phase.  We derive some of
the scaling properties of a particular model in this class which can
be related to directed percolation
\cite {directed percolation,domany-kinzel,kinzel}.
However, unlike directed percolation, the model has no absorbing states
(to be discussed below).
The scaling predictions are compared to Monte-Carlo simulation results.

\paragraph {Model Description:}
\label{sec:modeldescription}
The class of models is most simply introduced in the language of interface
growth \cite{KimKosterlitz}, in which both adsorption and
desorption processes take place. In the present
models, desorption may take place only at the edge of a plateau.
For concreteness, we study two particular models in this class, (a) a
restricted solid on solid (RSOS) version that may also be considered
in a charged particle representation,
and (b) an unrestricted model that may be related to directed percolation.
Both models are on a 1d lattice with periodic boundary conditions
and are defined as follows: Let $h_i$ be the (integer) height of the
interface at site $i$, $i=1\ldots N$. The interface evolves by choosing
a site $i$ at random and carrying out one of the two following processes:
(a) adsorption of an atom
\begin{equation}
\label{dyn1}
h_i \rightarrow h_{i}+1 \; \; \mbox {with probability } q
\end{equation}
and (b) a desorption of atoms from the edge of a step
\begin{eqnarray}
\label{dyn2}
h_i \rightarrow \mbox{min}(h_{i},h_{i+1}) \; \;
\mbox {with probability } (1-q)/2\\
\label{dyn3}
h_i \rightarrow \mbox{min}(h_{i},h_{i-1}) \; \;
\mbox {with probability } (1-q)/2
\end{eqnarray}
In the RSOS version, the restriction
\begin{equation}
|h_{i}-h_{i+1}| \leq 1
\end{equation}
is imposed at all sites. The RSOS version may be viewed as a driven
diffusion model of two oppositely charged types of particles. The charges
\begin{equation}
\label{ChargedRep}
c_{i,i+1} \;=\; h_{i+1}-h_i \;\in\; \{-1,0,+1\}
\end{equation}
are bond variables and
represent a change of height between adjacent interface sites (see Fig.1).
In this representation, the dynamical rules (\ref{dyn1})-(\ref{dyn3})
correspond to randomly selecting two neighboring bonds and
performing the following processes with probabilities as indicated
on the arrows:
%
%
%
\vspace{-2mm}
\begin{figure}[h]
\begin{eqnarray} \nonumber
\epsfxsize=78mm
\epsffile{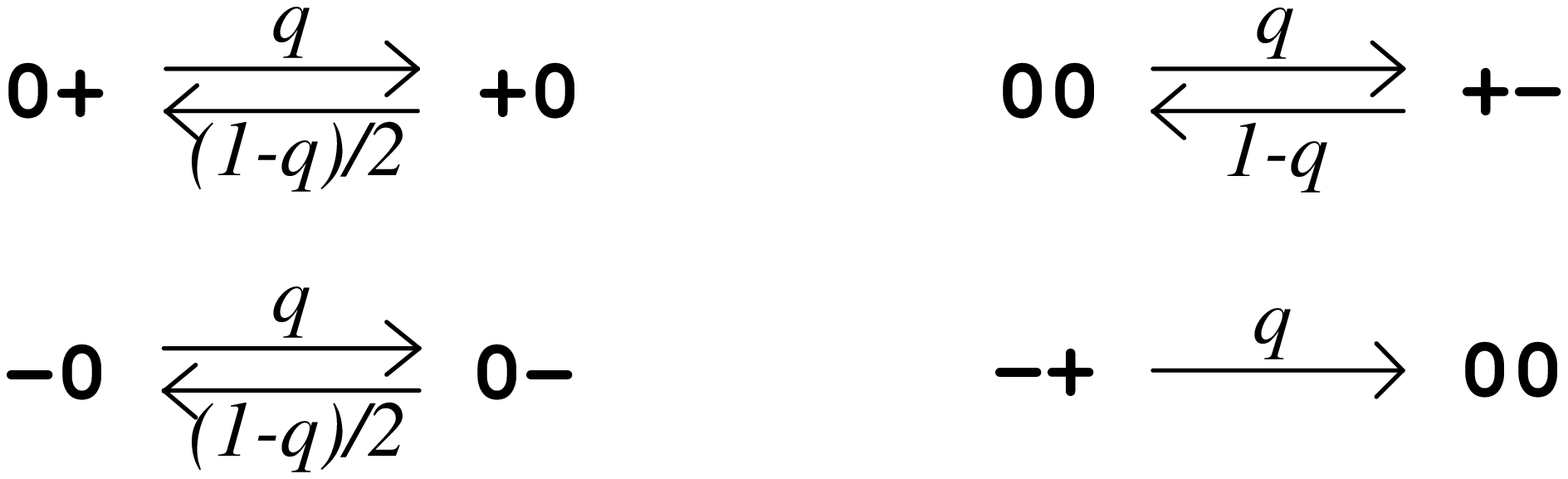}
\\[-27mm]
\label{ChargedRepProb}
\end{eqnarray}
\end{figure}
\noindent
\\ \indent
In both RSOS and unrestricted models,
when $q$ is small, a smooth phase is maintained.
In this phase the interface displays a small concentration of
short-lived islands, and its average velocity, $v$, is zero in the
 thermodynamic limit.
As $q$ increases, adsorption increases and typical islands
grow, until, above a critical value $q_c$, islands merge and
full new layers are completed, giving the interface a finite
growth velocity.
 Thus, when $q$ is small, {\em a local
mechanism that eliminates islands is present in the model }: an island
is formed with boundaries that are biased to move towards each other
(due to desorption from the island edges). The evolution of a large
island
\FigOne
for $q<q_c$ is illustrated in Fig. 2.
It is seen that the island shrinks,
ensuring the stability of the smooth phase.
This behavior is typical of islands of all sizes, except the
very largest (i.e. a complete layer).
The rule that no holes can be formed in a completed layer
($00 \not\rightarrow -+$ in the RSOS version), preventing it
from dividing into shrinking islands, is essential for
obtaining the smooth phase.
To demonstrate the existence of the roughening transition, we carried
out Monte Carlo simulations of both models. In this Letter we present some
of the results obtained in this study. A more detailed account will be
published elsewhere \cite{future}. The phase transition takes place at
$q_c=0.189 \pm 0.002$ for the RSOS model and $q_c=0.233 \pm 0.001$
for the unrestricted model. The interface width
is defined by the standard deviation of the height distribution
$w=[N^{-1} \sum_i (h_i - N^{-1}\sum_i h_i)^2]^{1/2}$.
We find that starting from a flat interface,
$w$ rises as
$w \sim t^{\chi/z}$ for short times, saturating for large $t$ at
$w \sim N^{\chi}$ where $N$
is the lattice size. At $q>q_c$, the numerical results are consistent with
the KPZ exponents \cite{KPZ} $\chi = \frac12$ and $z = \frac32$,
indicating a rough interface.
Below $q_c$, $w$ saturates at
a finite value independent of $N$, showing that the phase is smooth.
The critical behavior at $q=q_c$, shown in Fig.3, is
\begin{equation}
w \sim \log(N)
\end{equation}
In the following we discuss the symmetry breaking which takes place for
$q<q_c$. We also discuss the relation to directed percolation and
the critical behavior near $q_c$.

\paragraph{Spontaneous Symmetry Breaking:}
\label{sec:ssb}
To demonstrate some of the model's properties, it is convenient to
consider the RSOS model in the charged particle representation
(Eq. \ref{ChargedRepProb}, Fig. 1). At $q<q_c$,
the charges are arranged as closely bound $+-$ dipoles.
At $q>q_c$, the dipoles become unbound, and the fluctuations
in the total charge, measured over a distance of order $N$,
diverge with $N$.
Thus the transition is manifested in correlations between
\FigTwo
charged particles rather than in their density.
The symmetry breaking which takes place in this model is best seen by
introducing a coloring scheme by which
each of the sites between the charged particles is colored
in one of two colors, such that the two sites adjacent to
a $+$ or $-$ particle have different colors, and the two
sites adjacent to a $0$ particle have the same color (Fig.1b).
Every move in the dynamics
corresponds to a local rearrangement of the charged particles
and site coloring. Under the model dynamics, any configuration
of charges and coloring can evolve to any other allowed
configuration of charges and coloring. Thus the model
has no absorbing states. However, at $q<q_c$, a typical configuration
displays unequal concentrations of the two colors. As the system evolves
the configurations flip between majority colors in a time scale which
was found to
grow exponentially with the system size \cite{future}.
Thus, though the dynamical rules are symmetric with
respect to the site colors, the system spontaneously selects one
of two colors as a majority color, breaking the symmetry.
The system in the phase space of charge configurations and
colorings is ergodic at any finite size, but becomes non-ergodic
in the thermodynamic limit when $q<q_c$.
To quantify this symmetry breaking, we define a magnetization-like
order parameter (valid for both the RSOS and unrestricted models)
\begin{equation}
M\;=\; \frac{1}{N} \, \sum_{i=1}^N(-1)^{h_i}
\label {eq M def}
\end{equation}
which can be envisaged by considering the two colors as
`up' and `down' spins.
The order parameter is clearly not conserved by the dynamical rules.
In the smooth phase ($q<q_c$), $\langle M \rangle \ne 0$
in the thermodynamic limit. On the other hand, in the rough phase
$\langle M\rangle=0$.
Monte-Carlo simulations (Fig.3)
show that near the phase-transition, for both models,
\begin{equation}
\label {eq M}
\langle |M| \rangle  \sim \epsilon^{\theta}, \;\;\;\;
 \theta=0.55 \pm 0.05\,.
\end{equation}
where $\epsilon=q_c-q$.
%
%
%
%
\paragraph {Relation to directed percolation:}
\label{sec:scaling}
Some features of the {\em unrestricted} model may be related
to a directed percolation (DP) model
\cite{directed percolation,domany-kinzel,kinzel}, allowing a derivation of
several scaling properties. The occupation of the lowest exposed
level ($n_0$ in Fig.1a) corresponds to the
wet or percolating sites. The non percolating or dry sites are the
sites where higher levels are occupied. A wet region may become
dry both at its edges and bulk sites, while a dry region may
shrink only at the edges. This defines a contact process \cite{jensen},
 which is a sequential update version of a DP model \cite {domany-kinzel}.
The percolating phase corresponds to the
smooth phase in the model obtained at $q<q_c$.  Thus, the occupation of the
lowest level should vanish at the transition with the exponent $\beta$
 characterizing the critical behavior of the DP wet phase
\begin{equation}
\label{eq n0}
n_0 \sim \epsilon^{x_0}, \;\;\;\; x_0=\beta \approx 0.28
\end{equation}
The front velocity for $q>q_c$ may be related to the life-time of
typical wet islands below the percolation threshold. These islands
have a lifetime which diverges at the percolation threshold,  with the
critical exponent $\nu_{||}$ of the DP coherence time
\cite{directed percolation}.
 The time to complete a new layer is the time it
takes for its missing regions (percolation wet sites) to be covered
by adsorption (dry up). Thus the velocity $v$ is proportional to the
inverse of the island lifetime:
\begin{equation}
\label {eq v}
v \sim (-\epsilon)^y, \;\;\;\;  y=\nu_{||} \approx 1.73
\end{equation}
These exponents are in good agreement with the values measured
by Monte-Carlo simulations,
$x_0=0.29 \pm 0.03$ and $y=1.7 \pm 0.1$, as shown in Fig.3.
For the RSOS model, there is no direct mapping to DP, since layers grown
on top of a dry island affect its evolution.
Monte-Carlo simulations and diagonalization of the time-evolution operator
of small systems ($N<12$) \cite {kinzel,future}, suggest that both RSOS and
unrestricted models have the same exponents, and therefore
 belong to the same universality class.

We now present a simple scaling argument for the behavior of  $n_k$,
 the density of sites
at height $k$ above the lowest exposed level. Consider first the lowest
 exposed level $k=0$. According to the DP picture, there are two length
 scales in the problem: the average size of the dry islands,  $l_0$,
 which diverges at the
transition as $l_0 \sim \epsilon ^{-\beta}$  \cite {kinzel},
and the transverse correlation length
$\xi_{\perp}\sim \epsilon^{-\nu_{\perp}}$
 with $\nu_{\perp} \approx 1.10$ \cite{kinzel}.
 The two length scales are related, for a system of size $N$, by the finite
 size
 scaling relation $l_0 \sim \epsilon^{-\beta} f(N \epsilon^{\nu_{\perp}})$,
where $f$ is a function satisfying $f(s) \sim s^{\beta/\nu_{\perp}}$ for
$s \rightarrow 0$. Similarly, $n_0 \sim
\epsilon^{\beta} g(N \epsilon^{\nu_{\perp}})$
 with $g(s) \sim s^{-\beta/\nu_{\perp}}$ for  $s \rightarrow 0$.
At criticality, one therefore has
 $n_0 \sim \l_0^{-1} \sim N^{-\beta/\nu_{\perp}}$.
\FigThree
We now consider the
level $k=1$. One may view islands of sites at heights
$k \ge 2$ as growing on top
 of the dry islands of the $k=1$ level, whose typical size is $l_0$.
 Applying the same scaling relations and assuming that the system size
may be replaced by $l_0$, we find
 $n_1 \sim \l_1^{-1} \sim l_0^{-\beta/\nu_{\perp}}$,
where $l_1$ is the mean
 size of islands of sites with height $k \ge 2$.
Repeating this reasoning for
the next levels, one has $n_k \sim l_k^{-1}
\sim \l_{k-1}^{-\beta/\nu_{\perp}}$.
 One therefore obtains
\begin{equation}
\label {eq n1}
n_k \sim \epsilon^{x_k}, \;\;\;\;  x_k=\beta (\beta/\nu_{\perp})^k
\end{equation}
\noindent
Numerical simulations (see Fig. 3) yield $x_1 = 0.07 \pm 0.02$, as
 compared with $x_1 \approx 0.07$ obtained form Eq. \ref {eq n1}.
 More extensive numerical
 simulations are needed to demonstrate the validity of Eq. \ref {eq n1}
for this exponent and the exponents associated
with higher levels.
\paragraph{Discussion}
\label{sec:discussion}
A class of models that describe 1d growth processes with a roughening
transition are presented. These models provide an example of spontaneous
symmetry breaking that takes place with a {\em non-conserved} order parameter
in a {\em ring} geometry.
 The zero rate of desorption of an atom who's two neighbors are on the
 same level is crucial for obtaining a smooth phase, and thus a roughening
transition.  The present models can
be viewed as a hierarchy of DP-like processes, with the $k$-th echelon
DP process confined to the dry sites of the $k-1$-echelon DP process.
The critical exponents $y$ and $x_k$ are found to be related
 to DP exponents.
It would be interesting to find out whether other exponents, such as
 $\theta$ (Eq. \ref {eq M}), may also be related to the DP problem.
 The models are easily generalized
to higher dimensions, where the mapping to directed percolation and the
scaling arguments are expected to apply.
 It would be of interest to construct a
coarse-grained field-theory \cite{KPZ,cardy} that describes the present
 class of
models.

\noindent
{\em Acknowledgments:} We thank B. Derrida, E. Domany, G. Grinstein,
C. Jayaparkash, S. Sandow and D. Wolf for helpful discussions. This
work was supported by The Minerva Foundation, Munich.


\end{document}